\documentclass[twocolumn]{aastex701}
\usepackage{CJK}
\newcommand{\azv}{AzV~493}

\newcommand{\msun}{$M_\odot$}
\newcommand{\heii}{He\thinspace{\sc ii}}
\newcommand{\hei}{He\thinspace{\sc i}}
\newcommand{\kms}{km s$^{-1}$}

\shorttitle{Extreme Oe Variable AzV 493}
\shortauthors{Oey et al.}

\received{2026 March 6}
\revised{2026 June 15}
\accepted{2026 June 22}

\submitjournal{PASP}

\begin{document} 

\title{Constraints on Binarity for the Extreme Oe Variable Star AzV 493}

\correspondingauthor{M. S. Oey, msoey@umich.edu}

\begin{CJK*}{UTF8}{gbsn}
\author[0000-0002-5808-1320]{M. S. Oey (黄香莉)}
\affiliation{Astronomy Department, University of Michigan, 1085 South University Avenue, Ann Arbor, MI 48109, USA}
\email{msoey@umich.edu}

\author[0000-0001-7046-6517]{Irene Vargas-Salazar}
\affiliation{Astronomy Department, University of Michigan, 1085 South University Avenue, Ann Arbor, MI 48109, USA}
\email{ivargasa@umich.edu}

\author[0000-0002-2397-206X]{Edmund Hodges-Kluck}
\affiliation{NASA/GSFC, Code 662, Greenbelt, MD 20771, USA}
\email{edmund.hodges-kluck@nasa.gov}

\author[0000-0003-0521-473X]{Norberto Castro}
\affiliation{Leibniz-Institut für Astrophysik, An der Sternwarte, 16, D-14482, Potsdam, Germany}
\email{ncastro@aip.de}

\author[0000-0002-0548-8995]{Micha\l\ K. Szyma\'nski}
\affiliation{Astronomical Observatory, University of Warsaw, Al. Ujazdowskie 4, 00-478 Warszawa, Poland}
\email{msz@astrouw.edu.pl}

\author{Mario Mateo}
\affiliation{Astronomy Department, University of Michigan, 1085 South University Avenue, Ann Arbor, MI 48109, USA}
\email{mmateo@umich.edu}

\author[0000-0002-6718-9472]{Mathieu Renzo}
\affiliation{Department of Astronomy, University of Arizona, Tucson, Arizona 85721}
\email{mrenzo@arizona.edu}

\author[0000-0003-0696-2983]{Mark W. Suffak}
\affiliation{Department of Physics and Astronomy, Western University, London ON N6A 3K7, Canada}
\email{msuffak@uwo.ca}

\author[0000-0002-0870-6388]{Maxwell Moe}
\affiliation{Department of Physics and Astronomy, University of Wyoming, 1000 E. University Ave., Dept. 3905, Laramie, WY 82071, USA}
\email{mmoe2@uwyo.edu}

\begin{abstract}

The extreme Oe star \azv\ is known to show unusual photometric and spectroscopic variability that suggest the presence of an unseen companion in a highly eccentric and long-period (7.3 or 14.6-year) orbit.  We obtained a Chandra/ACIS observation near the putative periastron for the 7.3-year orbit 
to test for
transient X-ray emission that would confirm its binary nature.  Our data only place an upper limit to the X-ray luminosity of $L_X < 2.5\times 10^{33}\ \rm erg\ s^{-1}$ based on the 0.5 -- 8 keV flux limit.  Additionally, we obtained 4 new spectroscopic observations with the M2FS spectrograph at Magellan 
and 20 archive FLAMES/GIRAFFE and X-Shooter spectra from ESO/VLT
to further constrain the possibility of radial velocity (RV) variation.  
Statistical analysis of the RV measurements yields inconclusive results regarding the existence of variations.  We discuss possible mass limits for a potential companion, which may be a black hole, in
the event that the variations are real.  
The violet-to-red (V/R) Balmer ratio has also recently inverted, which may be a further indication of a companion.

\end{abstract}

%% \keywords{\uat{Galaxies}{573} --- \uat{Cosmology}{343} --- \uat{High Energy astrophysics}{739} --- \uat{Interstellar medium}{847} --- \uat{Stellar astronomy}{1583} --- \uat{Solar physics}{1476}}

\keywords{early-type stars --- Oe stars --- Be stars  --- high-mass X-ray binary stars --- circumstellar disks --- stellar pulsations --- interacting binary stars --- compact objects --- runaway stars --- variable stars --- Small Magellanic Cloud}

\section{Introduction}

Massive, interacting binary stars are the progenitors of a wide variety of energetic phenomena including high-mass X-ray binaries (HMXBs), gamma-ray bursters, 
ultraluminous X-ray sources
(ULXs), stripped-envelope core-collapse supernovae (SNe), and gravitational wave events. It is therefore a vital task to sort out which binary parameters lead to which outcomes. 
OBe stars are now recognized to be a population of massive stars that generally have spun up to near-critical rotation speeds through mass and angular momentum transfer of binary companions \citep[e.g.,][]{bodensteiner:20, vinciguerra:20, pols:91}.  For most of these, the companions have already exploded as SNe \citep[e.g.,][]{Phillips2024, Dallas2022, DorigoJones2020, Boubert2018}. Thus, for systems that remain bound, many OBe stars are expected to be in eccentric binaries \citep[e.g.,][]{brandt:95,renzo:19walk}, and evidence suggests that this is indeed the case \citep{VargasSalazar2025}. 

The star AzV 493 \citep{Azzopardi1975} in the SMC is an extraordinary Oe star that appears to be a highly eccentric binary system \citep{Oey2023}.
\citet{2016ApJ...819...55G} identified AzV 493 as the earliest SMC Oe star, based on emission not only in the Balmer lines, but also in He {\sc i} and He {\sc ii} $\lambda4686$; the latter feature is rarely observed in other Oe stars \citep{1974ApJ...193..113C}. AzV 493 is therefore also likely to be the earliest Oe star known, making this system an especially valuable empirical constraint on the massive binary parameter space and associated models.
\citet[][hereafter Paper~I]{Oey2023} carried out an extensive study of this object based on both photometric and spectroscopic time-series data.
They obtain stellar effective temperature $T_{\rm eff} = 42000$ K, corresponding to an O4--5 spectral type, luminosity $\log L/L_\odot = 5.83 \pm 0.15$, radius $R/R_\odot = 15 \pm 3$, and projected rotational velocity $v\sin i = 370 \pm 40\ \rm km\ s^{-1}$. Based on single-star evolutionary tracks by \citet{2011A&A...530A.115B}, they estimate $M/M_\odot = 50 \pm 9$. 

Paper~I presents the remarkable
$I$ and $V$-band light curves of AzV 493 from the OGLE survey \citep{Udalski2015}, together with 11 spectroscopic observations, mostly obtained from 2015 to 2021.  The light curve shows a long-term cycle of either 7.28 or 14.55 years, as well as short-timescale, semiperiodic variations on the order of $\sim$30 and 40 days. As shown by that work, the latter may correspond to stellar radial pulsations, as suggested by the corresponding cycle in $V-I$ and pulsation period. 
The long-term cycle as measured from the light curve minima at MJD 52303 and MJD 57626 corresponds to a period of of 5311 days (14.55 years), with possible evidence that the period is alternatively half of this value, or 7.28 years.  The optical emission-line spectra show a wide range of variability, with evidence of a thick, obscuring outer disk component together with a thinner, inner disk that dominates the emission-line spectrum and which is sometimes occulted by the outer component.

Paper~I proposes that the long-term cycle corresponds to the repeated ejection of the circumstellar decretion disk, which occurs at the eruption, quickly occulting part of the photosphere and then gradually dissipating, causing the observed flux to increase.  Such a process could be induced by the presence of an unseen companion with a highly eccentric orbit similar to that of heartbeat stars.  As discussed in Paper~I, the photometric properties of the light curve eruption do not correspond directly to the pattern of true heartbeat stars; however, if a companion is still responsible for the long-term variability, then the implied eccentricity nevertheless places this star in a class with similarities to heartbeat stars.
A periastron separation of $\sim2R$, where $R$ is the stellar radius, is necessary for this scenario, and thus
the 14.6 (7.3)-year period implies eccentricity $e \sim 0.95$ (0.93), and apastron $\sim 43$ (28) AU.  
The angle of inclination $i$ must be fairly high to generate disk obscuration, implying that the rotational velocity also must be much higher than that observed, i.e., $> 400\ \rm km\ s^{-1}$. This high value is a natural signature of accretion from binary mass transfer \citep[e.g.,][]{packet:81,renzo:21zeta}.
We also note that most heartbeat stars show tidally excited oscillations (TEOs) caused by the close periastron interaction \citep[e.g.,][]{Fuller2017}; it is unclear whether the oscillation in AzV 493 could be due to TEOs, which normally would dissipate on much shorter timescales. If they are not TEOs, then it is unclear what their origin could be.  A triple system origin is also a possibility.

The extreme eccentricity is consistent with an origin from the supernova or core collapse of the donor star, implying that the unseen companion could likely be a neutron star or black hole. Paper~I finds that AzV 493 appears most likely to be a runaway star with Gaia DR3 peculiar velocity of $54\pm 11\ \rm km\ s^{-1}$ and originating from the star-forming complex N84. 
This high velocity may suggest that the system also experienced dynamical acceleration from its birth cluster, but in any case, its high eccentricity and Oe star status still support a post-SN status.

Since AzV 493 is a rapidly rotating Oe star with a prominent decretion disk, binary mass transfer has almost certainly occurred. This has been invoked to explain low-mass compact objects in very young regions \citep{belczynski:08}, and in particular, with very massive companions \citep{vandermeij:21}, such as AzV 493. The average initial 
close
binary mass ratio for O stars is inferred to be $q = M_2/M_1 \sim 0.5$, with a relatively flat distribution \citep[e.g.,][]{Moe2017}. 
We adopt this as a representative value of $q$.
If mass transfer occurred during the donor's main sequence phase \citep[Case A;][]{kippenhahn:67}, it is expected to be slower and more mass-conserving, implying a zero-age-main-sequence (ZAMS) mass of $M_1 \sim 30 - 40$ \msun\ for the adopted $q$, also accounting for the final donor core mass. On the other hand, if mass transfer occurred after the donor main sequence (Case B), then it should take place rapidly, in which case
non-conservative mass loss is far more likely, implying an even higher ZAMS mass for $M_1$.
While these expectations may be oversimplified \citep[e.g.,][]{Wang2021,Lechien2025,Sen2025}, they suffice for providing first-order picture of the primordial system.

\citet{VargasSalazar2025} 
suggest that the unseen companion is likely a black hole based on the locus of high-probability contours of binary population synthesis models generated with BPASS \citep{Eldridge2017, Stanway2018}.
However, this interpretation is sensitive to the high mass of the observed Oe star, which may be overestimated if the envelope is inflated as has been suggested for OBe stars \citep[e.g.,][]{Castro2018}. If \azv\ has a substantially lower mass, then a neutron star companion has been thought to be more likely \citep[e.g.,][]{oconnor2011,sukhbold:16,zhang2008}.
However, recent work suggests that black holes may originate from a variety of core-collapse and supernova scenarios \citep{Burrows2025} that are not simply a function of mass, but are sensitive to multiple stellar physical parameters and metallicity.
We also caution that the data allow for the possibility that the companion is a normal pre-SN star with mass $< 7$ \msun, based on the spectroscopic detection limit \citep{VargasSalazar2025}, or a He star; for these scenarios, the high-eccentricity orbit would be more difficult to explain.  

Thus, establishing whether AzV 493 has a neutron star or other type of companion sets vital observational constraints on the nature and fate of high-mass binary systems and their compact remnants.
Confirming the nature of the companion to a star with such a high estimated ZAMS mass, and with a significant kick, would provide a cornerstone for theoretical models of the explosion and the binary interactions preceding it.
Given the high $v\sin i$ and varying spectral features of AzV 493 
(e.g., Figure \ref{fig:spectra} below), 
radial velocity (RV) monitoring has been difficult \citep[Paper~I;][]{VargasSalazar2025}.
In this work, we follow up on Paper~I with X-ray observations and additional radial velocity measurements to further refine constraints to the system and the nature of the likely companion.

\begin{figure*}[htp]
    \centering
     \includegraphics[width=1.3\textwidth,angle=90]{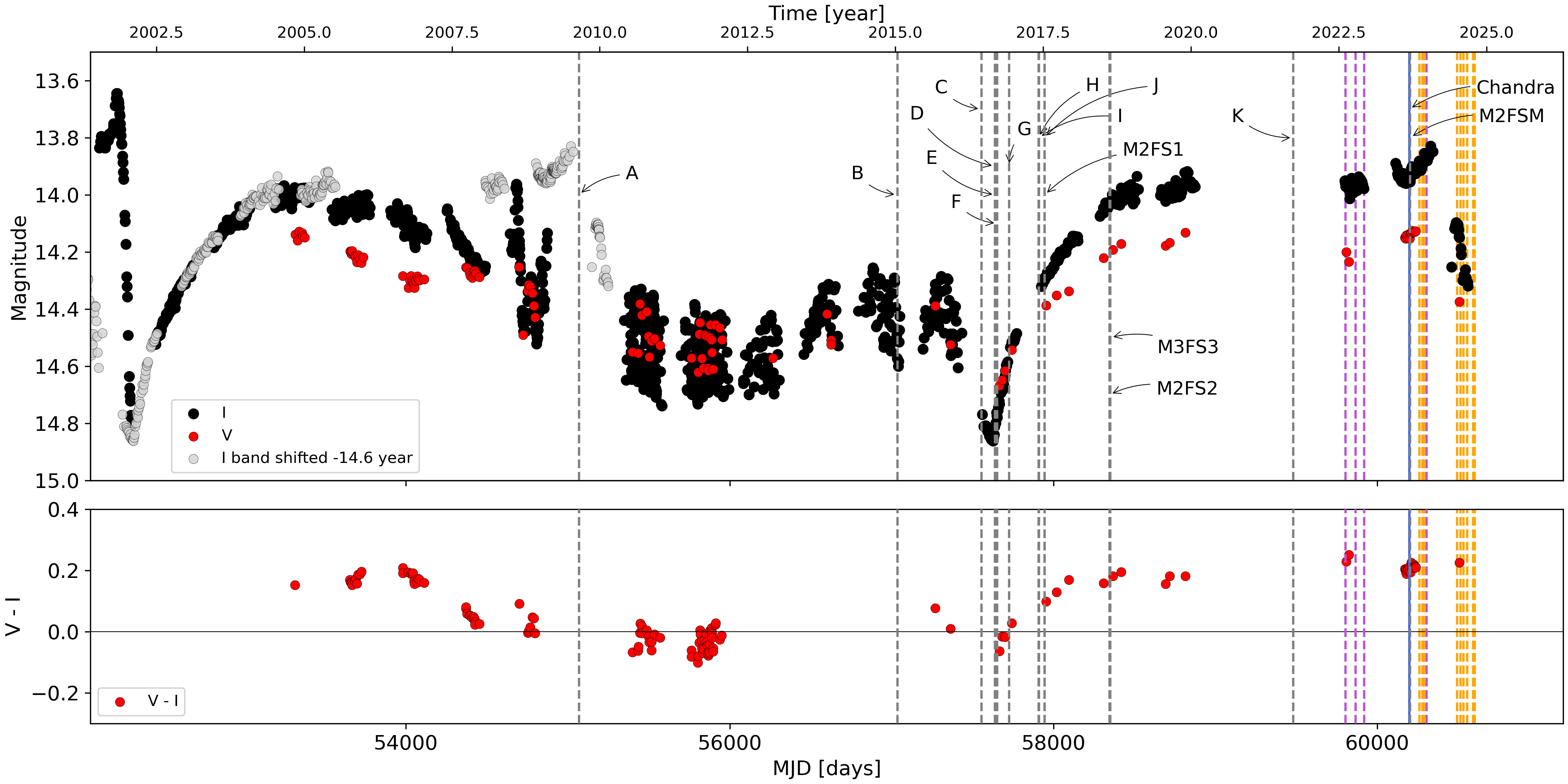}
    \caption{OGLE light curve in $I$ and $V$ shown in black and red, respectively.  Dates for our Magellan spectroscopic observations are shown with 
    black dashed lines, and those for X-Shooter data are shown in purple and FLAMES/GIRAFFE in orange.  The Chandra observation is indicated by the blue solid line. The gray points show the most recent photometry shifted back by the long-period phase of 14.6 years.
    The bottom panel shows the $V-I$ color variation.}
    \label{fig:phot}
\end{figure*}

\clearpage
\section{X-ray Data}\label{sec:Xray}

Detection of a hard X-ray signal would definitively confirm the existence of a compact object, 
either a neutron star or black hole.
AzV 493 is not known to be a high-mass X-ray binary (HMXB) in the SMC catalog of \citet{Antoniou2019}, and a 74.33 ks Chandra/HRC archive observation of the region in 2012 (ObsID 14054) shows no detection at the position of AzV 493. 
However, given the strong interaction at periastron, it is possible that a neutron star companion could temporarily acquire an accretion disk and thus become a transient HMXB. 
If this were the case for AzV 493, it would be by far the earliest and most extreme eccentricity OBeXRB known \citep{Brown2018}.

We therefore obtained Chandra ACIS-S observations of AzV 493 on 11 September 2023 (MJD 60198) (ObsID 26516; PI Oey).  
This date assumes that periastron occured on 2023 August $30 \pm 5$ days, based on the 7.3-year period.
The exposure was 19.40 ks with the target observed using the S3 chip. 

There is no detection of AzV 493, and we place a 3$\sigma$ upper limit of 5.9 counts using the CIAO {\tt srcflux} tool.
Assuming an absorbing column $N_H = 1.6\times 10^{22}\ \rm cm^{-2}$ 
and a power-law source SED with $\Gamma = 1.4$, typical for BeXRBs, we obtain
a corresponding 
0.5 -- 8 keV upper limit of $F_X < 5.3\times10^{-15}\ \rm erg\ s^{-1} cm^2$, or $L_X < 2.5\times 10^{33}\ \rm erg\ s^{-1}$ 
at the adopted SMC distance of 62 kpc \citep{Graczyk2014}. This is based on $A_V = 1.22$ for \azv\ \citep{Lamb2016} and taking the mean $N_{\rm HI}/A_V = 13.2\ \rm cm^{-2}$ for the SMC Bar from \citet{Gordon2003}.  Since the object is located in the SMC Wing, we also consider the ratio of  $N_{\rm HI}/A_V = 7.4\ \rm cm^{-2}$, which yields 
$L_X < 1.4 \times 10^{33}\ \rm erg\ s^{-1}$; however, we caution that this $N_{\rm HI}/A_V$ value is based on the extinction curve for only a single star \citep{Gordon2003}.
If we instead adopt a Wing distance of 55 kpc \citep{Nidever2013, Cignoni2009}, the derived $L_X$ limit decreases by $\sim$ 21\%.

Typical quiescent BeXRBs have $\log L_X/\rm erg\ s^{-1} \sim 33 - 34$ in the SMC \citep{Laycock2010}, while flares can reach $\log L_X/\rm erg\ s^{-1} \sim 37$ \citep{Brown2018}; thus, our non-detection suggests that \azv\ may not host a neutron star, which would be consistent with the scenario that the companion is a black hole as predicted by \citet{VargasSalazar2025}.  
However, Figure~\ref{fig:phot} shows the updated light curve using OGLE $I$ and $V$ photometry \citep{Udalski2015}.  We see that the flux of \azv\ continues to increase for approximately 130 days after the Chandra observation, thus it is also possible that the Chandra observation may have taken place too early to catch any new X-ray activity due to a neutron star.
Recent simulations by \citet{Rast2025} show that for highly eccentric orbits, $L_X$ in BeXRB systems is indeed cyclic and also depends on orbital parameters, with accretion rates that are much lower for strong misalignment between the orbital and rotational planes, and for retrograde orbits.  The apastron values can be 2 or more orders of magnitude lower than our detection limit (R. Rast, private communication).
A lower-mass companion such as a He star or white dwarf also cannot be ruled out.

Another possibility is that the system period corresponds to the longer value of 14.6 years.  Figure~\ref{fig:phot} shows that the light curve does not repeat itself exactly, and there is discrepancy around the long cycle's half-period around MJD 55000.  
It therefore remains unclear whether the observed date corresponds to an actual periastron. 

\section{Radial velocities}

In the course of the OB/e-star spectroscopic monitoring campaign of the SMC Wing by \citet{VargasSalazar2025}, we obtained 4 additional spectra of AzV 493 using the Michigan/Magellan Fiber System (M2FS), a multi-object fiber spectrograph at the 6.5-m Magellan Clay Telescope at Las Campanas Observatory.  Three of these were obtained with the same setup as the M2FS spectra in Paper~I (Epochs M2FS1 -- M2FS3), in particular covering the wavelength range $\sim 4080 - 4465$ \AA, which includes H$\gamma$, H$\delta$, and \heii\ $\lambda$4200.  In addition, we obtained a new spectrum near the time of the X-ray observations with M2FS using the MedRes mode (Epoch M2FSM), covering 8 echelle orders with total wavelength range $\sim 4140 - 5430$ \AA. 
There are also a number of archive spectra taken with FLAMES/GIRAFFE
(Program IDs 112.25R7.001, 112.25R7.002, and 112.25R7.003; PI T. Shenar)
and X-Shooter (Program IDs 109.22V0.001, PI: D. Pauli; and 112.25D1.001  PI: F. Tramper)
at the ESO/VLT 8-m telescopes during the period 2022 August to 2024 November, which spans the epoch of the Chandra observation.  The GIRAFFE data cover the wavelength range
3960--4560 \AA\ with $R\sim6700$, and the X-Shooter UVB-arm data cover 3200--5900 \AA\ with $R\sim6700$.
Table~\ref{tab:obs} gives an updated summary of the spectroscopic observations from this work and Paper~I; in addition to M2FS, the data in Paper~I were also obtained with the IMACS and MIKE instruments at Magellan.

Figure~\ref{fig:spectra} shows the rectified spectra. 
As noted, the epoch M2FSM was obtained near the time of the Chandra observation, and
just preceding the peak brightness.  It morphologically resembles Epoch~K, taken about two years earlier, and which displays stronger Balmer emission.  However, Epoch M2FSM lacks the  broad Balmer wings noted in Paper~I for Epoch~K. If this effect is real, it may
suggest that the new spectrum has a more obscured view of the innermost disk with the fastest orbiting material.
Figure~\ref{fig:spectra} also shows that the Balmer emission is strong in the recent period, continuing what was seen at Epoch~K; these Balmer levels have not been seen since Epoch~A, more than 10 years earlier, with the exception of the unusual spectrum of Epoch~F, which features an apparent infall event (Paper~I).  These new spectra are obtained during a photometrically bright period (Figure~\ref{fig:phot}).
Furthermore, the violet-to-red (V/R) ratio is now seen to vary.  In earlier epochs, the Balmer peaks are either similar in strength or stronger in the blue, but after roughly MJD 60500, 
during a period of declining luminosity, they begin to favor the red.
This supports the possibility that a companion at periastron was responsible \citep[e.g.,][]{Rast2024,Katahira2023}, although it is unclear whether the timing of the V/R change is consistent with the predicted periastron near the Chandra observation (2023 August 30; MJD 60186), which is about 400 days earlier. 

\begin{figure*}[htp]
    \centering
     \includegraphics[width=1\textwidth]{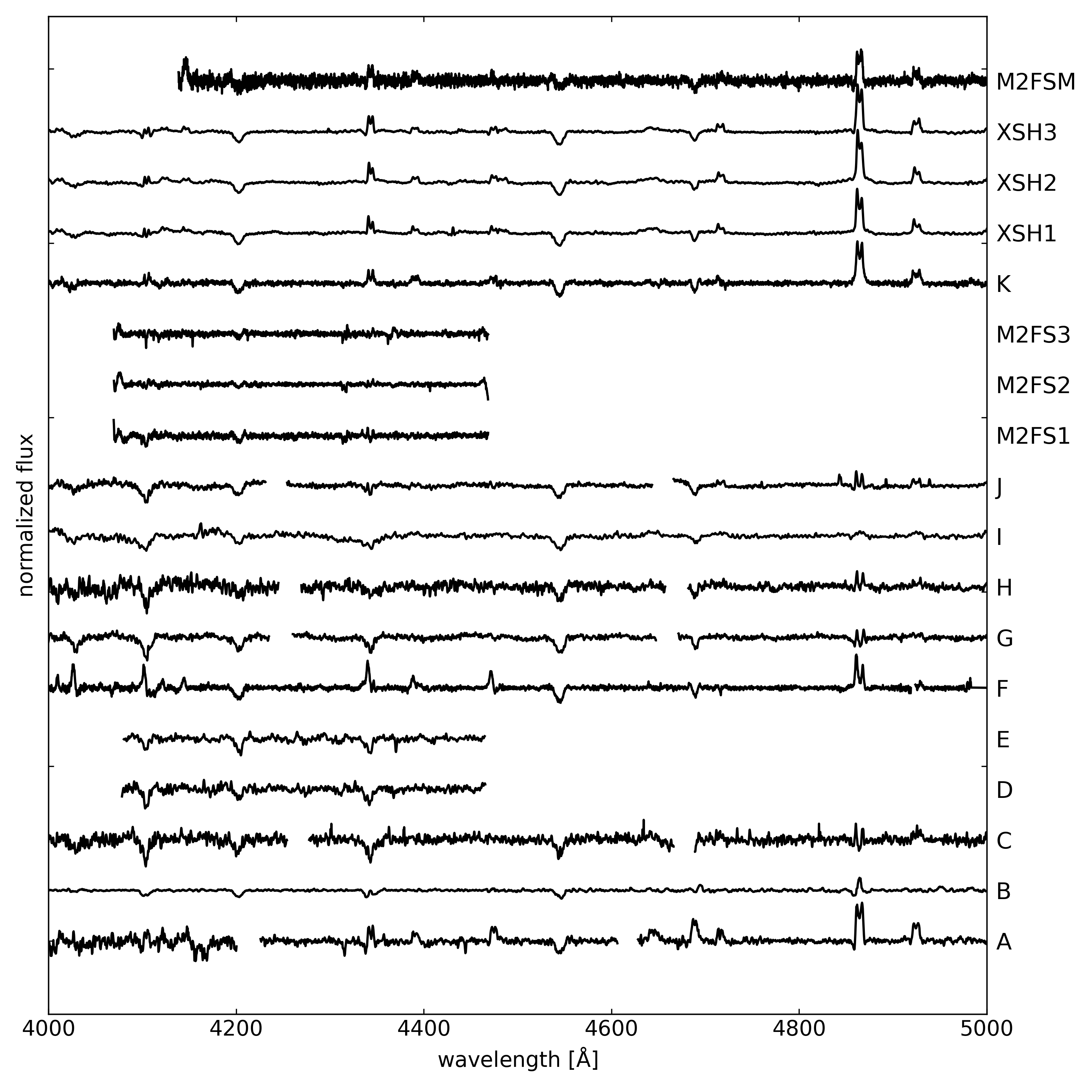}
      \caption{
     The AzV 493 multi-epoch spectroscopic observations sorted by increasing MJD from bottom to top, and normalized to the continuum. }
    \label{fig:spectra}
\end{figure*}

\setcounter{figure}{1}
\begin{figure*}[htp]
    \centering
     \includegraphics[width=1\textwidth]{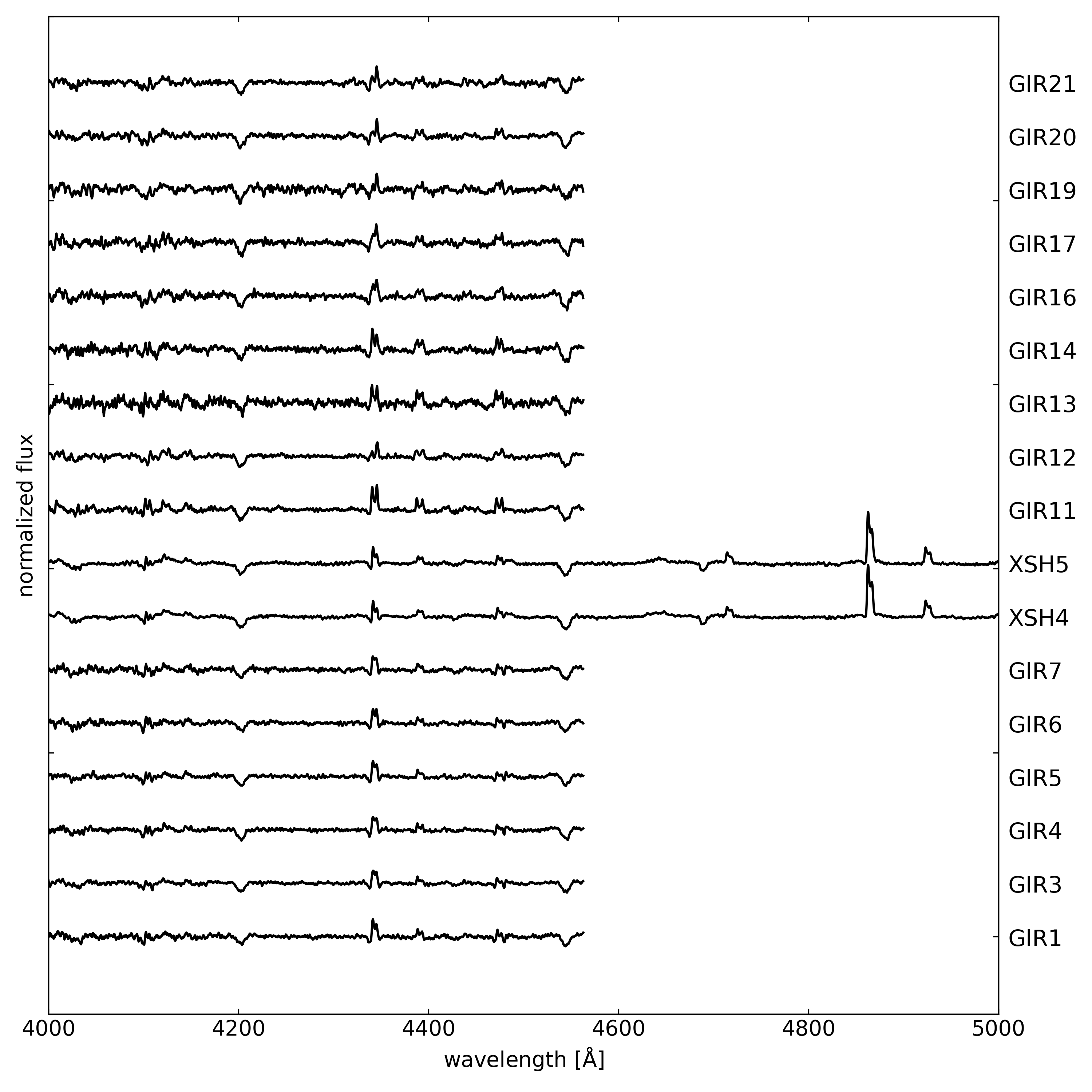}
      \caption{--- Continued.}
\end{figure*}

\noindent
\begin{table*}
	\caption{Spectroscopic Observations of \azv}
	\label{tab:obs}
	\begin{center}
		\begin{footnotesize}
			\begin{tabular}{cccrrcccl}
				\hline
			\hline
		Epoch & Date [UTC] &MJD  & S/N & $R$~~~ & Wavelength & Phase\tablenotemark{a} & 
        RV\tablenotemark{b} & Instrument  \\
		 &  & &  &  & Range [\AA] & & (\kms) &   \\
	\hline
A & 2009-08-26T01:43:36.0 & 55069.071944 & 140 & 3000 &3825--5422 & 0.076 (0.538)  & $152\pm 200$  & IMACS \\
B & 2015-01-14T02:12:03.0 & 57036.091701 & 120 & 28000 &3362--9397  &  0.817 (0.908) & $192\pm 18$ & MIKE \\
C & 2016-06-15T07:47:54.3 & 57554.324935 & 130 & 3000 &  3879--5479 & 0.012 (0.006) & $201\pm 60$ & IMACS \\
D & 2016-09-08T01:42:08.0 & 57639.070926 & 60 & 28000&  4079--4466 & 0.044 (0.022) & $217\pm 50$  &  M2FS \\
E & 2016-09-11T02:49:33.0 & 57642.117743 & 90 & 28000 &4080--4465 & 0.045 (0.022) & $239\pm 46$ &  M2FS \\
F & 2016-09-22T05:36:51.0 & 57653.233924 & 150 & 28000 & 3538--9397 & 0.049 (0.024) & $192\pm 29$ & MIKE \\
G & 2016-12-04T04:09:41.5 & 57726.173397 & 110 & 3000 &  3862--5458 & 0.076 (0.038) & $191\pm 38$ & IMACS \\
H & 2017-06-05T06:35:11.2 & 57909.274435 & 50 & 3000 & 3871--5471 & 0.145 (0.073) & $217\pm 55$ & IMACS \\
I & 2017-06-07T08:08:18.9 & 57911.339108 & 130 & 1300 & 3900--8000 & 0.146 (0.073) & $231\pm 83$ & IMACS\tablenotemark{c} \\
J & 2017-07-10T09:05:00.5 & 57944.378478 & 190 &3000 & 3854--5468 & 0.159 (0.079)& $187\pm 39$  &   IMACS \\
M2FS1 & 2017-07-11T2017-07-11 & 57945.29162  & 62 & 28000 & 4080--4465 & 0.159 (0.080) & $225\pm 36$ & M2FS\\
M2FS2 & 2018-08-16T2018-08-16 & 58346.31706  & 91 & 28000 & 4080--4465 & 0.31 (0.155) & $195\pm 28$ & M2FS\\
M2FS3 & 2018-08-20T2018-08-20 & 58350.35565  & 60 & 28000 & 4080--4465 & 0.312 (0.156) & $160\pm 40$ & M2FS\\
K & 2021-09-25T07:38:18.0 & 59482.318264 & 210 & 28000 & 3362--9397 & 0.738 (0.369) & $183\pm 17$  &  MIKE \\
XSH1 & 2022-08-13T06:43:16.0929 & 59804.280047 & 121 & 6700 & 3200--5900 & 0.859 (0.43) & $193\pm 17$ & XSHOOTER \\
XSH2 & 2022-10-14T02:42:44.003 & 59866.113009 & 143 & 6700 & 3200--5900 & 0.882 (0.441) & $189\pm 19$ & XSHOOTER \\
XSH3 & 2022-12-06T02:29:52.008 & 59919.104074 & 120 & 6700 & 3200--5900 & 0.902 (0.451) & $193\pm 16$ & XSHOOTER \\
M2FSM & 2023-09-16T08:58:18.0 & 60203.37382 & 60 & 27000 & 4138--5428 & 0.009 (0.050)   & $246\pm 53$ &  M2FS \\
GIR1 & 2023-11-13T03:38:11.022 & 60261.151516 & 74 & 6700 & 3959--4562 & 0.031 (0.516) & $192\pm 29$ & GIRAFFE \\
GIR3 & 2023-12-02T00:12:23.725 & 60280.008608 & 97 & 6700 & 3959--4562 & 0.038 (0.519) & $182\pm 23$ & GIRAFFE \\
GIR4 & 2023-12-04T02:43:32.081 & 60282.113566 & 79 & 6700 & 3959--4562 & 0.039 (0.52) & $191\pm 24$ & GIRAFFE \\
GIR5 & 2023-12-07T01:56:29.487 & 60285.080897 & 82 & 6700 & 3959--4562 & 0.04 (0.52) & $205\pm 24$ & GIRAFFE \\
GIR6 & 2023-12-08T04:59:51.666 & 60286.208237 & 75 & 6700 & 3959--4562 & 0.041 (0.52) & $183\pm 31$ & GIRAFFE \\
GIR7 & 2023-12-10T03:15:43.175 & 60288.135916 & 64 & 6700 & 3959--4562 & 0.041 (0.521) & $181\pm 32$ & GIRAFFE \\
XSH4 & 2023-12-26T02:35:55.1798 & 60304.108278 & 107 & 6700 & 3200--5900 & 0.047 (0.524) & $202\pm 21$ & XSHOOTER \\
XSH5 & 2023-12-26T02:57:30.421 & 60304.123269 & 93 & 6700 & 3200--5900 & 0.047 (0.524) & $192\pm 21$ & XSHOOTER \\
GIR11 & 2024-07-02T08:51:31.880 & 60493.369119 & 74 & 6700 & 3960--4562 & 0.119 (0.559) & $203\pm 27$ & GIRAFFE \\
GIR12 & 2024-07-24T08:34:12.020 & 60515.357084 & 78 & 6700 & 3960--4562 & 0.127 (0.563) & $203\pm 24$ & GIRAFFE \\
GIR13 & 2024-08-11T06:49:22.795 & 60533.284292 & 31 & 6700 & 3959--4562 & 0.134 (0.567) & $210\pm 57$ & GIRAFFE \\
GIR14 & 2024-09-04T03:47:32.292 & 60557.158013 & 46 & 6700 & 3959--4562 & 0.143 (0.571) & $182\pm 40$ & GIRAFFE \\
GIR16 & 2024-10-10T04:38:58.415 & 60593.193732 & 48 & 6700 & 3959--4562 & 0.156 (0.578) & $185\pm 38$ & GIRAFFE \\
GIR17 & 2024-10-12T01:24:51.858 & 60595.058934 & 55 & 6700 & 3959--4562 & 0.157 (0.578) & $208\pm 33$ & GIRAFFE \\
GIR19 & 2024-10-18T01:14:34.100 & 60601.051784 & 50 & 6700 & 3959--4562 & 0.159 (0.58) & $179\pm 46$ & GIRAFFE \\
GIR20 & 2024-10-19T02:24:41.301 & 60602.100478 & 66 & 6700 & 3959--4562 & 0.16 (0.58) & $190\pm 30$ & GIRAFFE \\
GIR21 & 2024-10-20T03:03:40.784 & 60603.127555 & 60 & 6700 & 3959--4562 & 0.16 (0.58) & $192\pm 33$ & GIRAFFE \\
\hline
\end{tabular}
\end{footnotesize}
\end{center}
\tablenotetext{a}{Phase relative to the light curve peak on MJD 52212, adopting a period of 2655.5 (5311) days. }
\tablenotetext{b}{RV measured by cross-correlation with PoWR model (see text).}
\tablenotemark{c}{Epoch I was observed with the f/2 camera while the other IMACS observations were obtained with the f/4 camera.}
\end{table*}

\begin{figure*}
    \centering
     \includegraphics[width=1.0\textwidth]{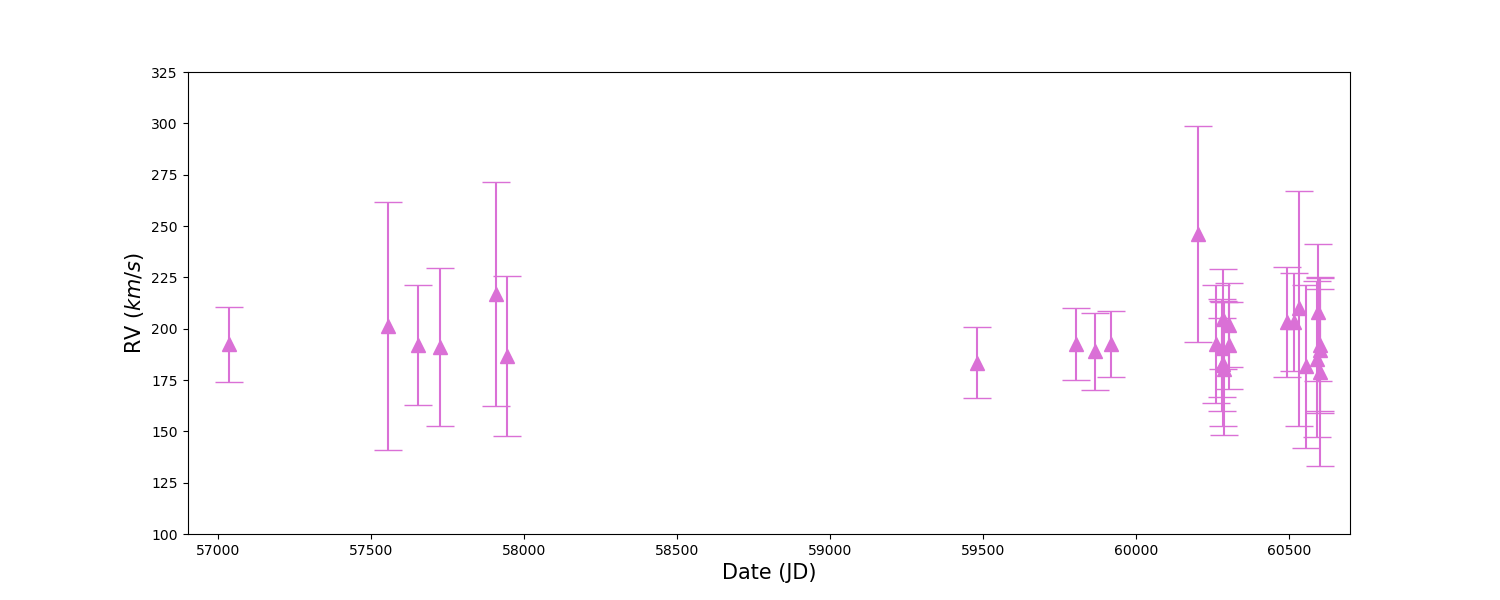}
     \includegraphics[width=0.45\textwidth]{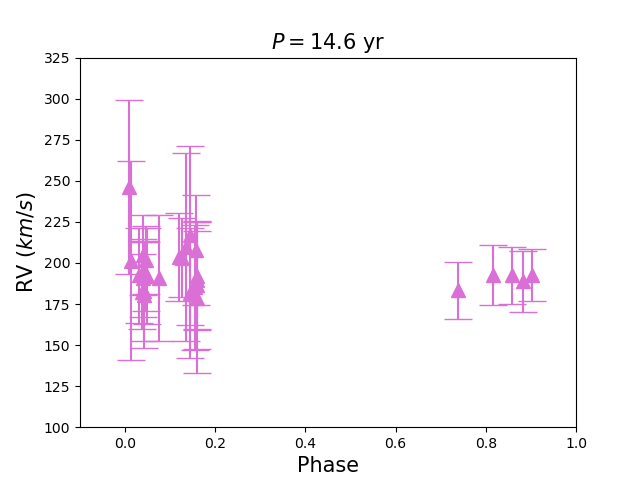}
     \includegraphics[width=0.45\textwidth]{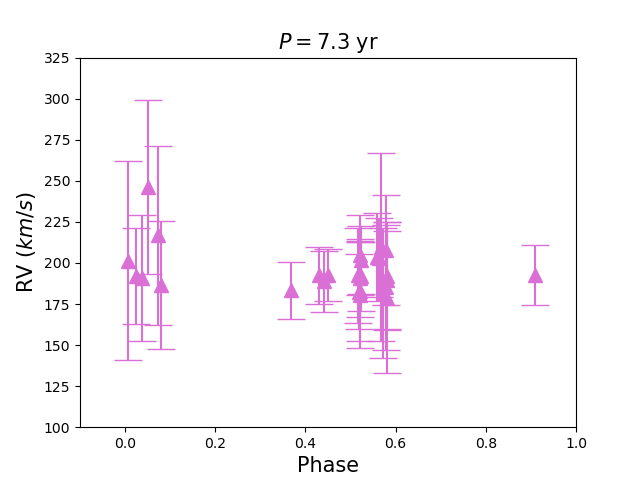}
    \caption{
    The top panel shows the radial velocity curve for \azv, and
    the bottom panels show the data in phase space assuming a 14.6-year (left) and 7.3-year period (right).
    Epochs~A, D, E, I, and M2FS1 -- M2FS3 are omitted from these data due to their large uncertainties (see text).  
    }
    \label{fig:rv}
\end{figure*}

As noted before, the high $v\sin i$ of AzV 493 and the emission in its \hei\ lines, typical of early Oe stars \citep{2016ApJ...819...55G}, has made measurement of radial velocities (RVs) especially difficult.  This star has by far the largest measurement uncertainties among objects in the SMC Wing sample of \citet{VargasSalazar2025}.  Here we reevaluate and remeasure the lines for all the spectra from Paper~I and \citet{VargasSalazar2025}, and we include new measurements for the new spectra in Table~\ref{tab:obs}.

We measure the RVs by Markov-chain Monte Carlo cross-correlation fitting with a template spectrum \citep{Becker2015}.  Following \citet{VargasSalazar2025}, we examine results based on
using a PoWR model \citep{Hainich2019} for $T_{\rm eff} = 40 - 42$ kK and $\log g/(\rm cm\ s^{-2}) = 2.0 - 4.4$ as the
template spectrum, and also a template corresponding to one of the observations, Epoch K, which has
the best signal-to-noise and representative features.  
This allows us to identify relative RV variations, but not the systemic velocities, which we obtain by averaging Gaussian fits to \heii\ 4200 \AA, \heii\ 4540 \AA, and \heii\ 4686 \AA\ absorption lines.
However, we find that the errors are systematically lower for the PoWR model template (median 32 \kms\ vs 42 \kms), and so we adopt these for the final RV measurements.
The spectra were smoothed and interpolated to
match the resolution of the IMACS spectra, which have the lowest resolution.  
Since the H and He I lines tend to show emission in Oe stars, we avoid using them for the cross-correlation fits.  Thus,
for the M2FS spectra and epochs F, G, H, I and J, we fitted only the \heii\ 4200 \AA, \heii\ 4540 \AA, and \heii\ 4686 \AA\ absorption lines.  Epochs B and C additionally mask the line \heii\ 4686 \AA, since it also shows emission. For 
Epochs D, E, and M2FS1 -- M2FS3, 
the only \heii\ line available is 4200 \AA, due to the limited spectral range.
Our updated RV measurements are given in Table~\ref{tab:obs}.
Epochs A, D, E, I, and M2FS1 -- M2FS3 all have measurement errors $>70$ \kms\ using either the model and/or the data template fits, which is more than twice the median error for the model fits.  We therefore drop these 7 data points to create a culled sample that we consider to be the most reliable dataset.
The RV curve for the culled sample is shown in Figure~\ref{fig:rv} in both MJD and phase.

\begin{figure*}
    \centering
     \includegraphics[width=0.45\textwidth]{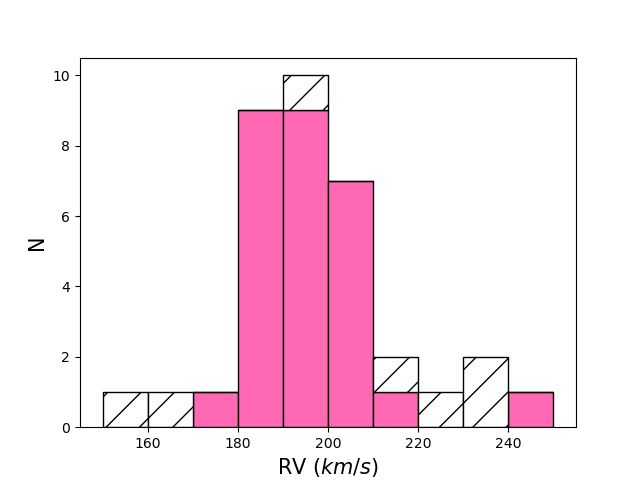}\quad
     \includegraphics[width=0.45\textwidth]
         {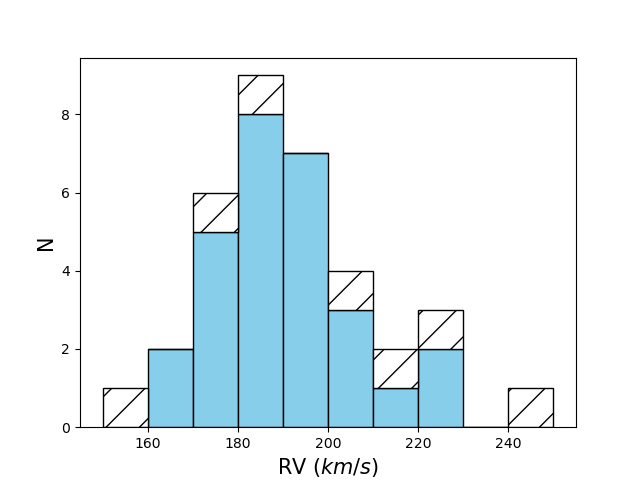}
    \caption{
    Radial velocity distributions for \azv\ based on the PoWR model template (left) and Epoch~K template (right).  The contributions of the high-uncertainty Epochs~A, D, E, I, and M2FS1 -- M2FS3 are marked by the hatched bars.}
    \label{fig:RVdist}
\end{figure*}

We use these revised RV measurements to reevaluate whether RV variability due to a companion object is detected.  If the object has a constant RV and the variation in our values is only due to random measurement errors, then we should expect them to have a normal distribution.  The Anderson-Darling and Shapiro-Wilk tests both evaluate normality in a single distribution \citep{RazaliYap2011}, and
we apply these tests to our sample of RV measurements.
Table~\ref{t_normality} shows the normality test results for both the entire sample and the culled sample, as well as the median, mean, and standard deviation.  Figure~\ref{fig:RVdist} shows the RV distributions, indicating the culled data points.
For good measure, we also show the distribution for the measurements based on the empirical template, which is broader, consistent with the larger errors in that sample.

While the plotted distribution looks fairly gaussian in appearance, Table~\ref{t_normality} shows that the statistical tests for the model-template measurements have $p$-values that are all at or below $\sim0.05$ for both the culled and total datasets, indicating significant deviation from a normal distribution.
The tests for the empirical-template measurements do not show statistically significant deviation from normality, perhaps due to their larger errors.
The Anderson-Darling test is more sensitive to the shape of the tails of the distribution, while the Shapiro-Wilk test is more sensitive to outliers and is also considered to be better for small samples.  However, the fact that the culled sample removed proportionately more outliers may bias the outcome of the Shapiro-Wilk test, where we see that the $p$-values are significantly lower for the culled sample. 

The statistical finding of a non-normal distribution for the RV measurements does not necessarily mean that RV variability is present, however.  
The tests assume that the measurements are all independent, but they originate from different instruments with different systematic uncertainties, which may generate relationships between subsets of the observations.  This could potentially generate a non-gaussian distribution for the combined sample.  Table~\ref{t_normality} also shows results for the 15 GIRAFFE data and the remaining non-GIRAFFE data, confirming significant differences in the standard deviations of these subsets.  Thus, the combined dataset may not correspond to a normal distribution even for random sampling of a single value; the same argument applies to the non-GIRAFFE data, which are themselves heterogenous.
We therefore regard these results as potentially suggestive, but inconclusive on their own.
% Note that this is a sum of gaussian distributions, rather than a set of variables that sum two gaussian-distributed components.  Therefore the sum is not gaussian.

\begin{deluxetable}{lccccc} \label{t_normality}
\tablecaption{Normality tests}
\tablehead{ & \colhead{Anderson-Darling}& \colhead{Shapiro-Wilk} & \colhead{Median} & 
\colhead{Mean} & \colhead{Std dev}\\
 & \colhead{$p$-value} & \colhead{$p$-value}  & \colhead{ \kms} & \colhead{\kms }& \colhead{\kms }
} \startdata
Culled data\tablenotemark{a} & 0.010 & 0.0003 & 192 & 195 & 14 \\
All data & 0.012 & 0.055 & 192 & 197 & 19 \\
GIRAFFE data & 0.095 & 0.088 & 191 & 192 & 10 \\
non-GIRAFFE\tablenotemark{a} & 0.010 & 0.0007 & 192 & 198 & 16 \\
\enddata
\tablenotetext{a}{Omitting Epochs A, D, E, I, and M2FS1 -- M2FS3.}
\end{deluxetable}

As demonstrated in Paper~I, the nature of the photometric and spectroscopic variability of \azv\ does strongly suggest the existence of an unseen binary companion.  This is further supported by the likely Oe-star history of binary mass transfer.   If the observed RV variations are real, we can use them to reevaluate the constraints on the putative companion mass, taking a revised maximum RV semiamplitude variation
from the culled sample of 34 \kms.
Based on the findings in Paper~I, we adopt eccentricity $e$ in the range 0.80 -- 0.98, inclination $i = 45^\circ - 90^\circ$, and current primary star mass $M_0$ of 50 \msun.  Following \citet{VargasSalazar2025}, we obtain lower limits on the companion mass of 
6\msun\ and 
8\msun\ 
for orbital periods $P$ of 7.3 and 14.6 years, respectively.  These values result from the extreme values for the eccentricity and inclination in our considered ranges, $e = 0.98$ and $i=90^\circ$, respectively, which we consider to be improbable.  For $e=0.8$, $i=60^\circ$, and $P = 7.3$ yr, the predicted companion masses increase to 
28\ \msun\ and 38\ \msun,
respectively,
implying a black hole, since the maximum-mass main-sequence star that would remain photospherically undetected is 7 \msun\ \citep{VargasSalazar2025}.
If we allow for the possibility of an inflated envelope and adopt a reduced mass for the current primary of $M_0 = 25$ \msun, then the corresponding lower limits for the companion mass are 
% 22 \msun and 31 \msun, 
19 \msun\ and 25 \msun,
respectively.  Even adopting the most extreme orbital parameters above, the companion mass is $>3$\msun\ for $P=7.3$ years. 
However, we reiterate that the dispersion in RV may not correspond to physical variation, in which case the mass of any secondary could be below these values, allowing for the possibility of a neutron star companion, 
or no companion.

\section{Conclusion}

In summary, we have obtained X-ray data and new spectroscopic data of the extreme Oe variable star \azv, in an attempt to confirm a binary nature for this object.  A neutron star companion could explain the general patterns seen in its spectroscopic and photometric variability, which suggest extreme orbital eccentricity, potentially attributable in part to a SN explosion.  

We place an upper limit of $L_X < 4.2\times 10^{33}\ \rm erg\ s^{-1}$ for our Chandra/ACIS observation near the time of the periastron for a putative 7.3-year orbit.  This constraint is on the low end for possible quiescent BeXRBs, suggesting that a neutron star companion is unlikely, but not ruled out.  Additional X-ray observations near the periastron of the alternative 14.6-yr orbit in 2030 would be of further interest.

We obtained 4 new Magellan/M2FS spectra, 
including one near the putative 7.3-year periastron.  Together with 15 FLAMES/GIRAFFE and 5 X-Shooter VLT spectra archived in recent years, we consider a total of 35
radial velocity observations when combined with the data in Paper~I.  
The V/R ratio has inverted from blue to red since 2024 October, which may be evidence for the existence of a binary companion.
The RV distribution statistically suggests that the measurements deviate from a normal distribution, supporting the possibility of orbital variation, although the heterogeneous nature of the dataset may invalidate these tests.
Improved methods for RV measurements could reduce the errors and also account for other issues such as potential non-radial pulsations; the center-of-gravity method described by \citet{Aerts2010} is a good example.

We obtain a kinematic lower limit for a potential companion of $\sim 6$\msun\ if the observed RV variations are real; if they are not, then an unobserved lower-mass companion may be present, or \azv\ may be a single star despite its compelling variability and likely binary history.

\begin{acknowledgments}

We thank Rina Rast for sharing model $L_X$ calculations.
We also thank the anonymous referee for suggestions that significantly improved this paper.

The scientific results reported in this article are based in part on observations made by the Chandra X-ray Observatory, ObsID 26516; https://doi.org/10.25574/26516.
This paper also includes data gathered with the 6.5-meter Magellan Telescopes and the OGLE project, both located at Las Campanas Observatory, Chile; 
and observations collected at the European Organisation for Astronomical Research in the Southern Hemisphere under ESO programs 109.22V0.001,276, and 112.25D1.001.
I.V-S. and M.S.O. acknowledge 
support for this work from the National Aeronautics and Space Administration through Chandra Award Number GO3-24020X, issued by the Chandra X-ray Center, which is operated by the Smithsonian Astrophysical Observatory for, and on behalf of, the National Aeronautics Space Administration under contract NAS8-03060.
N.C. acknowledges funding from the Deutsche Forschungsgemeinschaft (DFG) – CA 2551/1-2.
M.K.Sz. acknowledges support for the OGLE project from the Polish National
Science Centre grant OPUS-28 2024/55/B/ST9/00447.
We acknowledge publication support from the University of Michigan via the Big Ten Academic Alliance.
This research has made use of software provided by the Chandra X-ray Center in the application package CIAO. 

\end{acknowledgments}
 
% \vspace*{5mm}

\facilities{CXO, Magellan:Baade, Magellan:Clay, OGLE, VLT:Kueyen}

\software{CIAO \citep{fruscione2006},
scipy.stats \citep{2020SciPy},
astropy \citep{astropy:2013,astropy:2018,astropy:2022},
numpy \citep{numpy_harris2020array}
}

% \appendix
% \textcolor{blue}{
% \section{ASAS-SN photometry}
% }

% The ASAS-SN*** survey provides a large database of $V$-band photometry.  Here, we provide plots to examine the light curve and color variability with higher temporal resolution than is provided by OGLE-III data (Paper~I).  There is a significant, 0.19-mag* offset between the ASAS-SN data and the OGLE-III $V$-band data.  This can be attributed to the much larger *** arcsec pixel size of the ASAS-SN instrument, which therefore includes the light of many SMC background stars.  We therefore correct the ASAS-SN values with this median offset in the figures presented here, such that the corrected $V_* \equiv V_{\rm ASAS} + 0.19$.

% Figure~\ref{fig:ASASzoom} shows ****

\bibliographystyle{aasjournalv7}
\bibliography{Oey_AzV493_II}



\end{CJK*}
\end{document}